\documentclass[fleqn,usenatbib]{mnras}

\usepackage[T1]{fontenc}
\usepackage{newtxtext,newtxmath}
\usepackage{graphicx}
\usepackage{subcaption}
\usepackage{amsmath}
\usepackage{booktabs}
\usepackage{tabularx}
\usepackage{multirow}
\usepackage{xcolor}
\usepackage{microtype}

\hypersetup{
  hypertexnames=false,
  pdftitle={EON-SII: Design of a transportable picosecond stellar intensity interferometer for compact-star astrophysics},
  pdfauthor={Thomas Schweizer, Josef Eder, Razmik Mirzoyan, Carina Haider, Juergen Besenrieder, Olaf Reimann, Derek Strom, Roland Walter},
  pdfsubject={Reference design and projected performance of the EON-SII stellar intensity-interferometry pathfinder},
  pdfkeywords={stellar intensity interferometry, optical interferometry, telescope design, white dwarfs}
}

\title[EON-SII: 4~m diameter telescopes]{EON-SII: Design of a transportable picosecond stellar intensity interferometer for compact-star astrophysics}

\author[T. Schweizer et al.]{
Thomas Schweizer$^{1}$\thanks{E-mail: thomas.schweizer@mpp.mpg.de},
Josef Eder$^{1}$,
J\"urgen Besenrieder$^{1}$,
Razmik Mirzoyan$^{1}$,
Carina Haider$^{1}$,
\newauthor
Olaf Reimann$^{1}$,
Derek Strom$^{1}$,
Roland Walter$^{2}$
\\
$^{1}$Max Planck Institute for Physics, Boltzmannstrasse 8, 85748 Garching, Germany\\
$^{2}$Department of Astronomy, University of Geneva, Chemin Pegasi 51, 1290 Versoix, Switzerland
}

\date{Accepted XXX. Received YYY; in original form ZZZ}
\pubyear{2026}

\begin{document}

\label{firstpage}
\pagerange{\pageref{firstpage}--\pageref{lastpage}}
\maketitle

\begin{abstract}
Stellar intensity interferometry (SII) measures correlations in photon-arrival fluctuations recorded by telescopes observing bright celestial sources. It can resolve angular scales far smaller than those accessible to a single optical telescope and is largely insensitive to atmospheric turbulence. After the first demonstration of SII on Sirius in 1956, Hanbury Brown and Twiss used the technique to measure the diameters of 32 stars. More recently, VERITAS, MAGIC, H.E.S.S., and CTAO’s LST-1 have revived the method, although observations remain restricted to bright targets because of their optical design, optimized for  gamma-ray astrophysics, rather than SII.

We present EON-SII, the design and performance of a two-telescope intensity interferometer 
intended to extend the SII technique to compact targets at magnitudes of about
V=8.5 up to V=10.7. Each transportable telescope has a 4-m diameter mirror,
approximately $9\,\mathrm{m^2}$  collecting area, an actively aligned
18-panel primary mirror, and Cassegrain optics specified to concentrate at
least 90~\% of the light within 3 arcsec. A fibre-free spectrograph covers
400--550 nm at $R\simeq7000$--8000 and is designed to provide of order
$1000$ statistically independent spectral channels.

\end{abstract}


\section{Introduction}
\label{sec:introduction}

Stellar intensity interferometry was introduced by Robert Hanbury Brown and
Richard Twiss to measure stellar angular diameters from correlations between
intensity fluctuations recorded at separated telescopes
\citep{hbt1956,hbt1958}. The Narrabri Stellar Intensity Interferometer
subsequently measured the diameters of 32 stars and nine multiple-star systems
\citep{hanburybrown1974}. The method later received less attention than
amplitude interferometry, which offers higher sensitivity but requires coherent
beam combination and accurate optical-path control. Advances in
photon-counting detectors, timing electronics and digital correlators have now
renewed interest in intensity interferometry
\citep{lebohec2006,dravins2012,dravins2016}.

Modern SII systems have demonstrated this principle on arrays of atmospheric
Cherenkov telescopes. VERITAS recovered stellar diameters with four collectors
\citep{veritas2020}; MAGIC established a routinely operated two-telescope system
\citep{acciari2020magic,abe2024magic}; H.E.S.S. has reported first and two-colour measurements;
and joint MAGIC--LST-1 operation expands the collecting area and baseline set
\citep{hess2023,hess2025twocolour,magiclst2025}. Image
synthesis and photon-counting extensions are being developed in parallel
\citep{spolon2024,guerin2025photoncounting}. These facilities provide large
mirrors at modest incremental cost, but their optical point-spread functions,
focal-plane cameras and nanosecond-scale signal chains were optimized for air
showers. They therefore cannot simultaneously realize the compact focal spot,
high spectral channel splitting and picosecond timing required to maximize SII
sensitivity. The QUASAR programme follows the complementary route of
picosecond, multi-channel SII instruments on existing optical telescopes
\citep{walter2025quasar}.

\begin{figure}
\centering
\includegraphics[width=0.8\columnwidth]{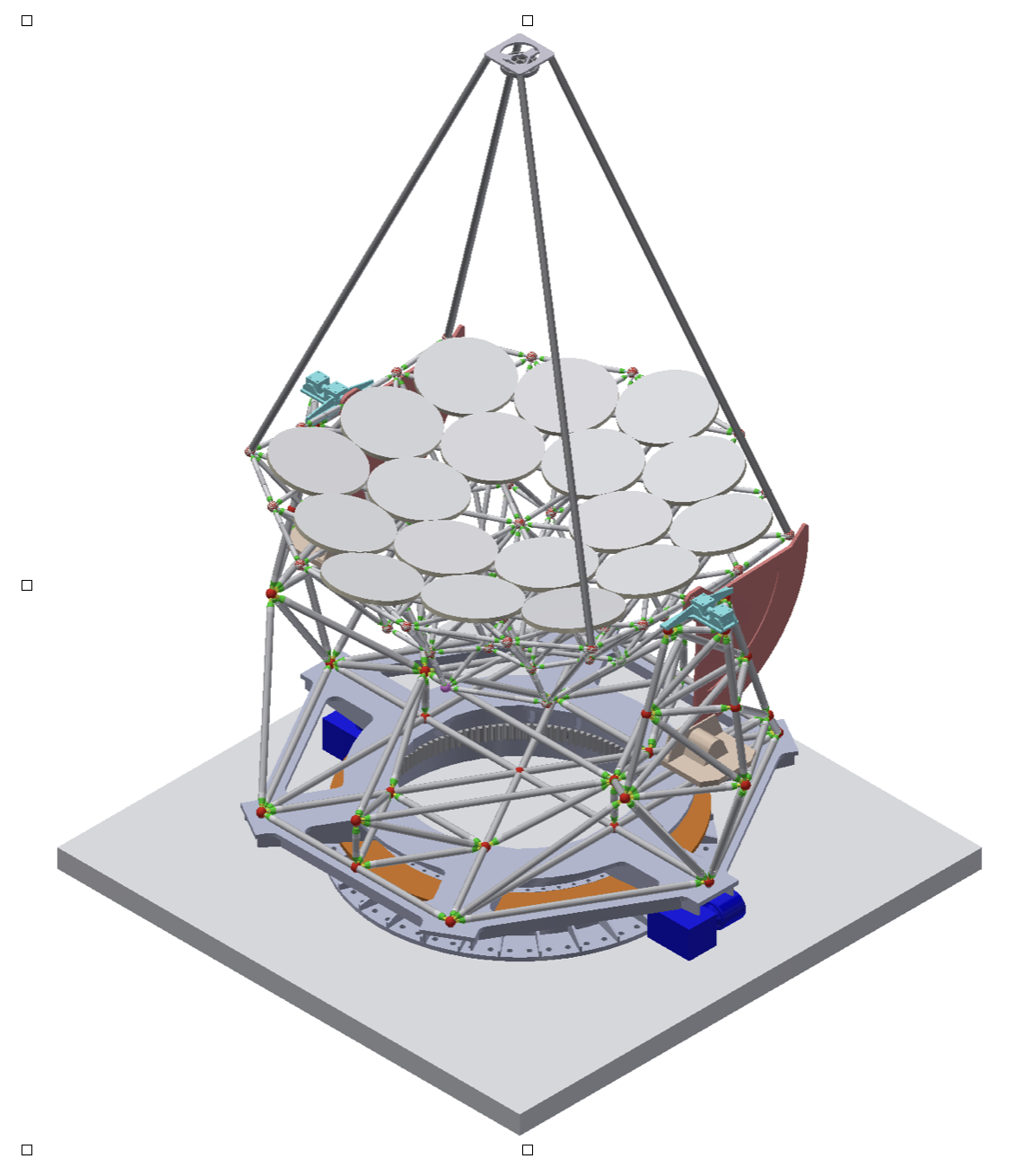}
\caption{Baseline mechanical design of the transportable 4-m EON-SII telescope
structure. }
\label{fig:telescope_design_concept}
\end{figure}

The telescope system proposed here (Fig.~\ref{fig:telescope_design_concept})
instead optimizes the complete telescope
optics, spectrograph and readout chain for photon-correlation sensitivity. It
combines high photon throughput from mirror to detector with approximately
1000 parallel spectral channels and 12--30 ps timing resolution. Each spectral
channel acts as an independent correlation measurement, so that spectral
multiplexing can increase the combined significance while preserving the
coherence contrast in each channel.

Since the sensitivity of an intensity interferometer is closely linked to the
angular size of the target, baseline reconfiguration is a key aspect of the
design. Transportable telescopes allow the baseline length to be adapted to the
visibility scale of different sources. The two-telescope system measures
visibility amplitudes and characteristic angular scales; Earth rotation changes
the projected baseline and provides additional sampling during an observation.

This paper presents the telescope design, the laboratory evidence for critical
components, and the projected performance of an end-to-end Monte Carlo model.

\section{Science motivation}
\label{sec:science_motivation}

A telescope design that reaches compact sources at $V\simeq8.5$--10.7 with
15--30 $\mu$as angular resolution in the 400nm-550nm wavelength range 
would open a largely unoccupied optical
observational regime. Existing optical interferometers are limited to angular
scales of order a few $10^2\,\mu$as, while many of the most interesting compact
stellar systems are an order of magnitude smaller and several magnitudes
fainter. Larger single-aperture telescopes cannot access this regime, because
their resolution is set by aperture diameter rather than kilometre-scale
baseline.

The key scientific opportunity is to measure physical sizes, not only fluxes.
For white dwarfs, angular diameters combined with parallaxes give geometrical
radii. Together with dynamical or gravitational-redshift masses, these radii
directly test the white-dwarf mass--radius relation and therefore the equation
of state of degenerate matter. Reaching $V\simeq8.5$--10.7 is important because
the nearby benchmark white dwarfs, including Sirius B, 40 Eridani B and
Procyon B, lie in this magnitude range and have expected angular diameters of
only a few $10\,\mu$as.

The same resolution and sensitivity add a spatial dimension to accretion
physics. In cataclysmic variables, novae and bright X-ray binaries, optical
light curves and spectra show when material brightens, but not directly where
the emission arises. Measuring the wavelength- and time-dependent half-light
radius $R_{1/2}(\lambda,t)$ during outbursts would distinguish compact inner-disc
heating from extended stream-impact or outflow emission, and would trace nova
ejecta expansion in continuum and diagnostic lines.

Thus the interest is not only higher resolution, but access to a new population:
faint, compact stellar systems whose angular sizes are tens of microarcseconds.
A dedicated intensity-interferometry telescope optimized for photon collection,
picosecond timing and many simultaneous spectral channels can make this regime
observable and define a scalable architecture for optical microarcsecond
imaging arrays.

\section{Measurement principle and telescope requirements}
\label{sec:principle}

\subsection{Observable and angular scale}

\begin{figure}
\centering
\includegraphics[width=\columnwidth]{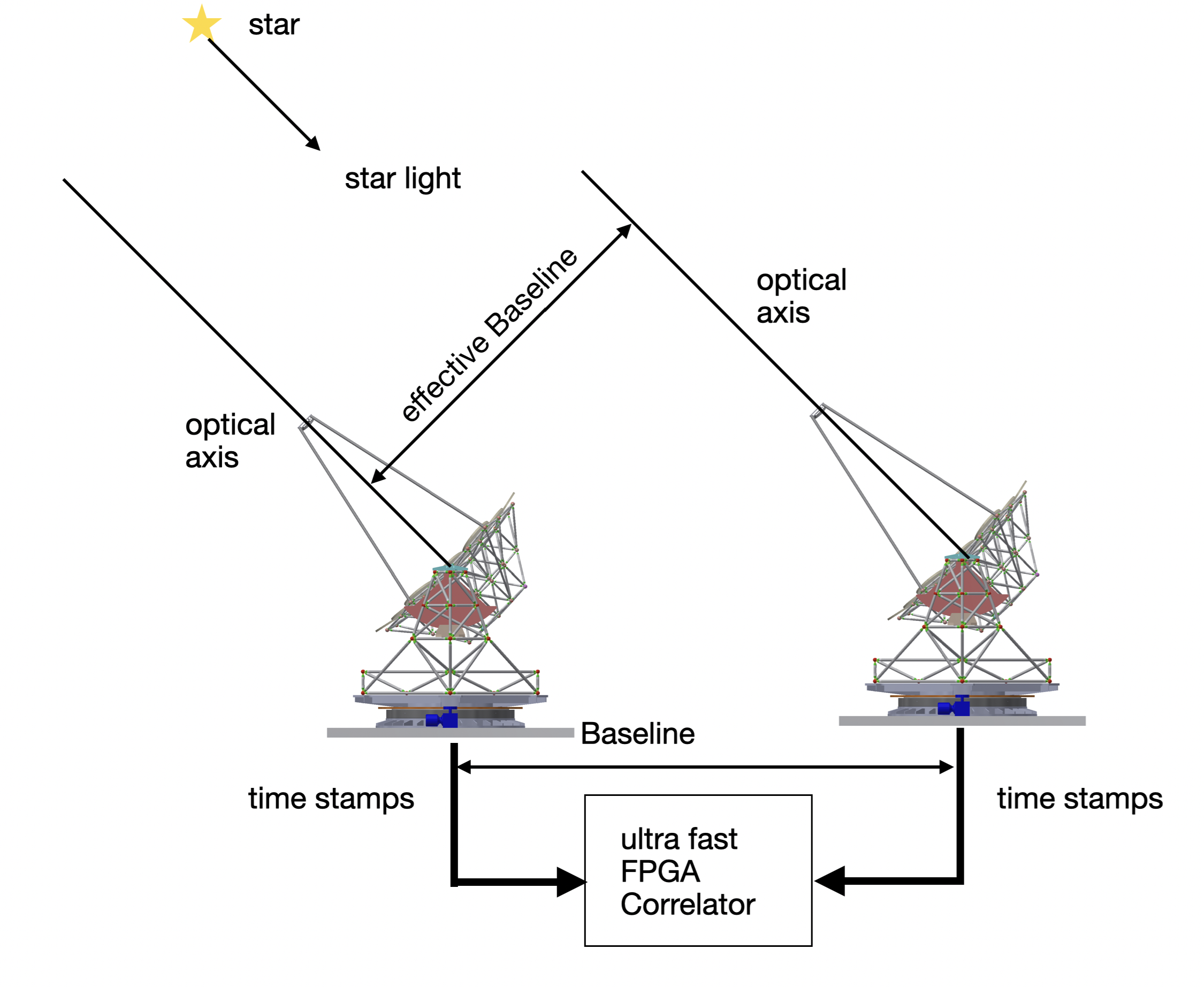}
\caption{Local photon time-tagging and digital correlation concept from the
final B1 document. Each telescope records photon streams locally; the streams
are delay-corrected and correlated electronically.}
\label{fig:telescope_setup}
\end{figure}

For quasi-monochromatic chaotic light observed by telescopes 1 and 2, the
normalized second-order temporal coherence may be written as the product of a spatial term and a time-dependent term:

\begin{equation} \label{eq:second_order_coherence}
   g_{12}^{(2)}(\mathbf{B},\tau)
   =
   1+
   |\gamma_{12}(\mathbf{B})|^2\,|g^{(1)}(\tau)|^2.
\end{equation}

\(\gamma_{12}(\mathbf{B})\) is a first-order spatial coherence quantity.
Where $\mathbf{B}=\mathbf{r}_2-\mathbf{r}_1$ is the baseline vector between the two telescopes and $\tau$ is the coherence times scale.

Intensity interferometry measures its squared modulus through a second-order correlation.
This is the chaotic-light Siegert relation in the notation used for stellar
intensity interferometry \cite{glauber1963opticalcoherence,foellmi2009g2astrophysics,guerin2025photoncounting}.

By the van Cittert–Zernike theorem, the complex degree of spatial
coherence $\gamma_{12}$ is the normalised Fourier transform of the source
brightness distribution. The spatial frequencies are measured in the
u-v space by the baseline vector $\mathbf{B}$.

SII measures $|V(u,v)|^2=|\gamma_{12}|^2$.

For a uniform disc of angular diameter $\theta$, the monochromatic model is
\begin{equation}
|V|^2
=
\left[
\frac{2J_1(x)}{x}
\right]^2,
\qquad
x=\frac{\pi B\theta}{\lambda},
\label{eq:uniform_disc}
\end{equation}
where $J_1$ is the first-order Bessel
function. At 500 nm, $\lambda/B=68.8\,\mu$as for a 1.5-km baseline.
Such a baseline partially resolves a 30-$\mu$as source and samples a steep part
of its visibility curve; configurations approaching 3 km probe lower
visibilities and provide leverage on diameter and non-circular structure.


\begin{figure}
\centering
\includegraphics[width=0.8\columnwidth]{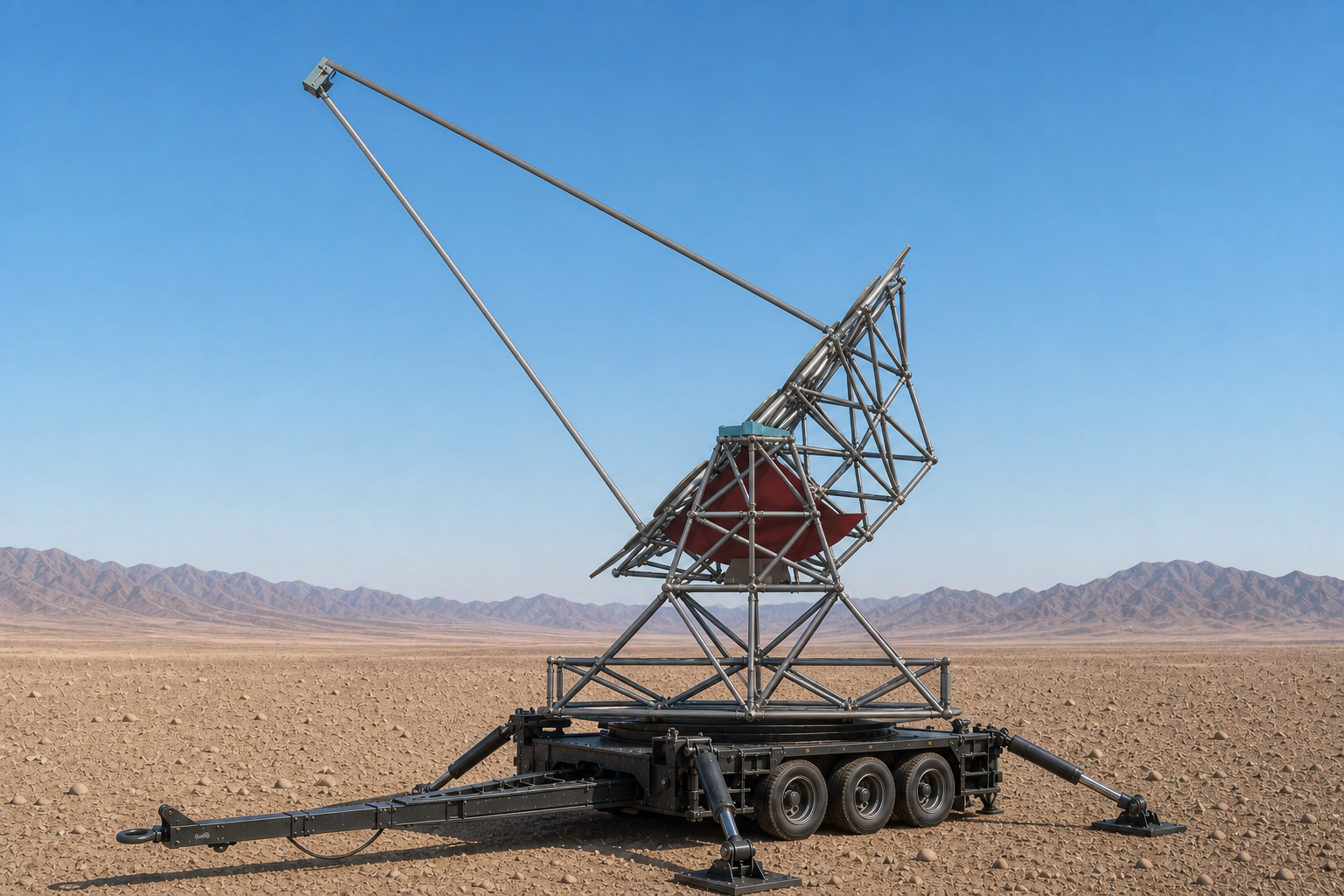}
\caption{The artistic image illustrates telescope deployment and transport.}
\label{fig:operation_concept}
\end{figure}

\subsection{Photon statistics and spectral wavelength splitting}

For a single spectral channel, the signal-to-noise ratio can be approximated by
the formula given by Hanbury Brown and Twiss

\begin{equation}
\left(\frac{S}{N}\right)_i
\simeq
A\,\alpha_i q_i n_{\nu,i}\,|V_i|^2
\frac{\sigma_{{\rm spec},i}}{F_i}
\sqrt{\frac{b_{{\rm el},i}T}{2}},
\label{eq:snr_analytic}
\end{equation}

where $A$ is collecting area, $\alpha_i$ detector quantum efficiency, $q_i$
optical throughput, $n_{\nu,i}$ source photon spectral density, $b_{{\rm el},i}$
effective electronic bandwidth, $T$ integration time, $F_i$ an excess-noise
factor, and $\sigma_{{\rm spec},i}$ a spectral-response efficiency factor.
The expression shows the central design requirements: collecting
area, detected photon efficiency and detector timing.

Splitting a broad optical band into channels improves the sensitivity
by statistically combining correlation measurements
\citep{trippe2014,tolila2024}.
For uncorrelated channels, the optimum combined significance is

\begin{equation}
\left(\frac{S}{N}\right)_{\rm tot}^2
=
\sum_{i=1}^{N_{\rm ch}}
\left(\frac{S}{N}\right)_i^2,
\label{eq:snr_channels}
\end{equation}

which approaches $\sqrt{N_{\rm ch}}$ times the single-channel significance when
channel responses and photon rates are comparable. Cross-talk, an
under-resolved line-spread function or common-mode electronic noise reduces the
effective number of independent channels and must be measured.

The corresponding photon-level estimator used in the design simulation counts
correlated and accidental pairs. In simplified form,

\begin{equation}
\frac{S}{N}
=
\frac{
\eta_{\rm vis} R_kR_l\tau_c |\gamma_{12}|^2T
}{
\left[
\eta_{\rm vis}R_kR_l\tau_c|\gamma_{12}|^2T
+2R_kR_l\Delta t_{\rm res}T
\right]^{1/2}},
\label{eq:snr_detected}
\end{equation}
where $R_k$ and $R_l$ are detected rates, $\tau_c$ is coherence time,
$\Delta t_{\rm res}$ is the effective coincidence resolution, and
$\eta_{\rm vis}$ describes contrast retention. This form makes explicit why
picosecond timing and narrow spectral channels are coupled requirements.

\subsection{EON-SII Design requirements}

Simulations show\cite{schweizer2026simulation} that the target sensitivity of magnitude 8.5-10.7 can
be reached by using telescopes with a minimum of $9~m^{2}$ mirror area,
1000 spectral channels and a timing resolution of 12ps-30ps FWHM. 

Table \ref{tab:requirements} lists the design requirements for such a two
telescope intensity interferometer.

\begin{table*}
\centering
\caption{Reference requirements for the proposed telescope design. Values are
current design targets unless explicitly identified as laboratory results.}
\label{tab:requirements}
\begin{tabularx}{\textwidth}{@{}p{0.22\textwidth}p{0.35\textwidth}X@{}}
\toprule
Subsystem & Reference value & Comment \\
\midrule
Telescope structure &
Two 4-m diameter mirrors; $\simeq9\,\mathrm{m^2}$ geometric area each &
Transportable structure with segmented mirror tiles. \\

Telescope baseline &
Reconfigurable baseline; nominal range 1.5--3 km &
Matches object sizes of approximately 20--70 $\mu$as in the visible band. \\

Drive control precision &
Drive pointing accuracy better than 3--5 arcsec &
Residual drive imperfections are corrected by the active mirror-control system. \\\

Focal concentration &
$\geq90$~\% of light within 3~as ($\simeq0.9$ mm) &
Enables fibre-free, efficient and stable spectrograph injection without
requiring diffraction-limited optics. \\

Active mirror-control precision &
The active mirror control shall position the PSF to better than 0.3 arcsec &
The focal spot must remain well centred within the entrance diaphragm of the spectrometer. \\

Spectral band &
Sensor sensitivity over 400--550 nm &
Optimized for white-dwarf spectra and available detector response. \\

Spectrograph &
$R\simeq7000$--8000; $\sim1000$ effective channels; $>60$~\% downstream throughput &
The resolving power must be high enough to provide about 1000 spectral channels
with spillover below 5~\%. \\

Combined mirror reflectivity &
80~\% in the performance model &
Maximizes photon throughput to the detector. \\

Isochronicity of the optics & must be better than 3~ps & The alignment of the
primary panels must keep isochronicity below 3~ps or 1~mm deviation in focal 
alignment. \\

Detector timing &
12--30 ps target &
Narrows the accidental-coincidence window and maximizes sensitivity. \\

Time stamping &
1024 channels with 3.125-ps nominal CERN picoTDC binning &
Must sample faster than the detector time resolution while remaining scalable
and cost-effective. \\

Sirius A rejection &
Leakage into the Sirius B channel $\lesssim10^{-5}$ &
Keeps the residual bright-star halo below 10~\% of the Sirius B photon
rate. \\

Replication constraint &
Hardware cost below EUR 1 million per telescope &
Constrains the architecture toward repeatable, moderate-cost array elements. \\
\bottomrule
\end{tabularx}
\end{table*}

\section{Telescope structure, primary mirror and active optics}
\label{sec:telescope}

The mount uses a lightweight steel space-frame architecture derived from the
structural principles of the MAGIC telescopes and the 23-m LST. The design
objective is not to suppress elastic deformation by means of a massive, highly
stiff structure, but to make deformation repeatable, observable and correctable
while maintaining safe eigenfrequencies, acceptable drive loads and robust wind
response. This approach requires precision metrology of the mirror pointing
directions and actuators that compensate structural deformation. The same
principle has proven effective in the MAGIC and LST telescope designs
\citep{schweizer2026mainstructure}.

Isochronicity is a key requirement for the optical design. Since the photon streams from the two telescopes are correlated on picosecond time-scales, the telescope optics must not dominate the timing response. We therefore require the optical path-length spread introduced by each telescope to be below 3 ps. This translates into a mechanical tolerance of approximately 1 mm on the dish deformation, the panel fixation, and the mirror-surface figure relative to the prescribed parabolic shape.

\begin{figure}
    \centering
    \begin{subfigure}[t]{0.4\linewidth}
        \centering
        \includegraphics[width=\linewidth]{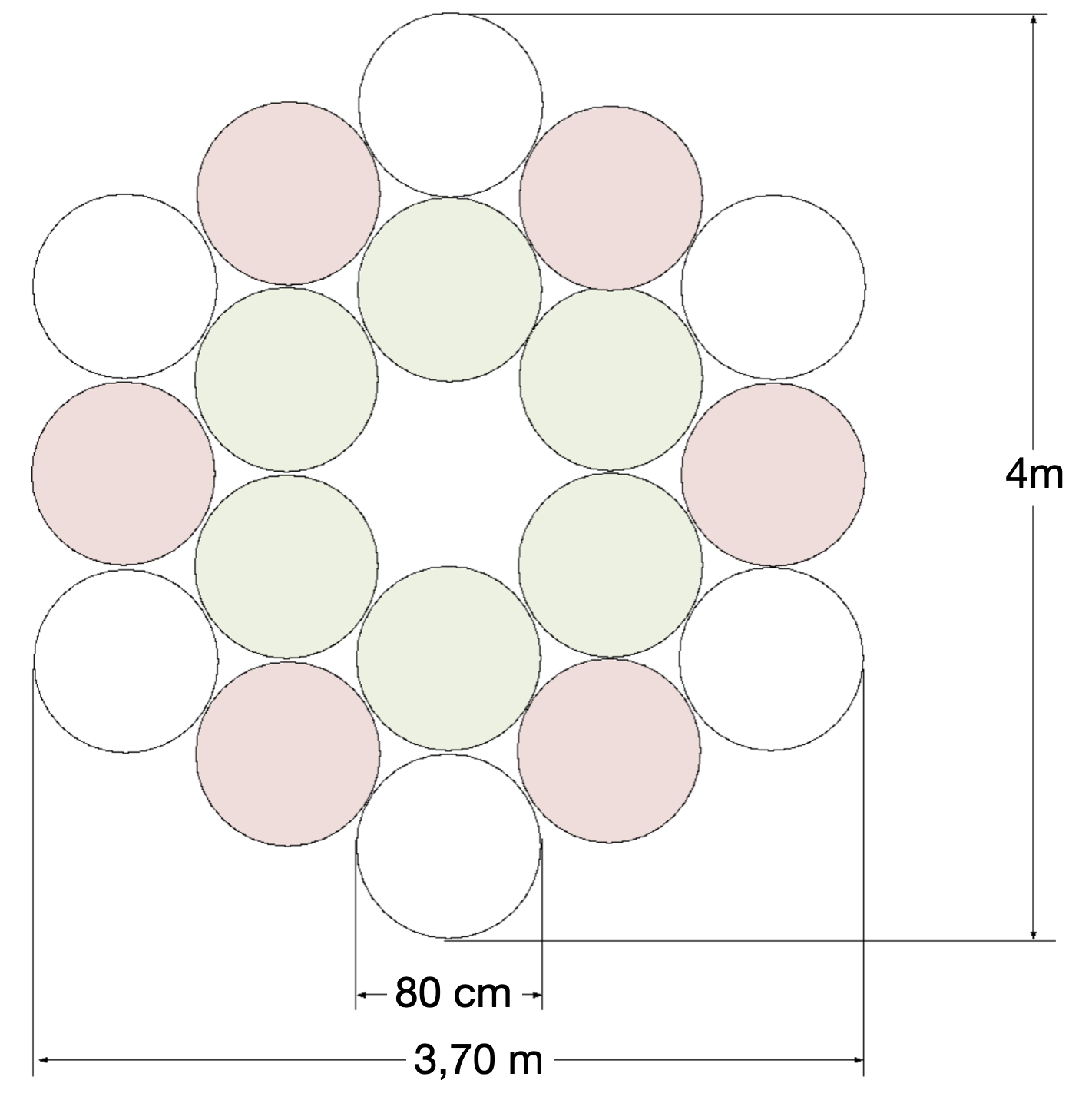}
        \caption{Segmented mirror layout.}
        \label{fig:round_mirror_design}
    \end{subfigure}\hfill
    \centering
    \begin{subfigure}[t]{0.58\linewidth}
    \centering
    \includegraphics[width=\linewidth]{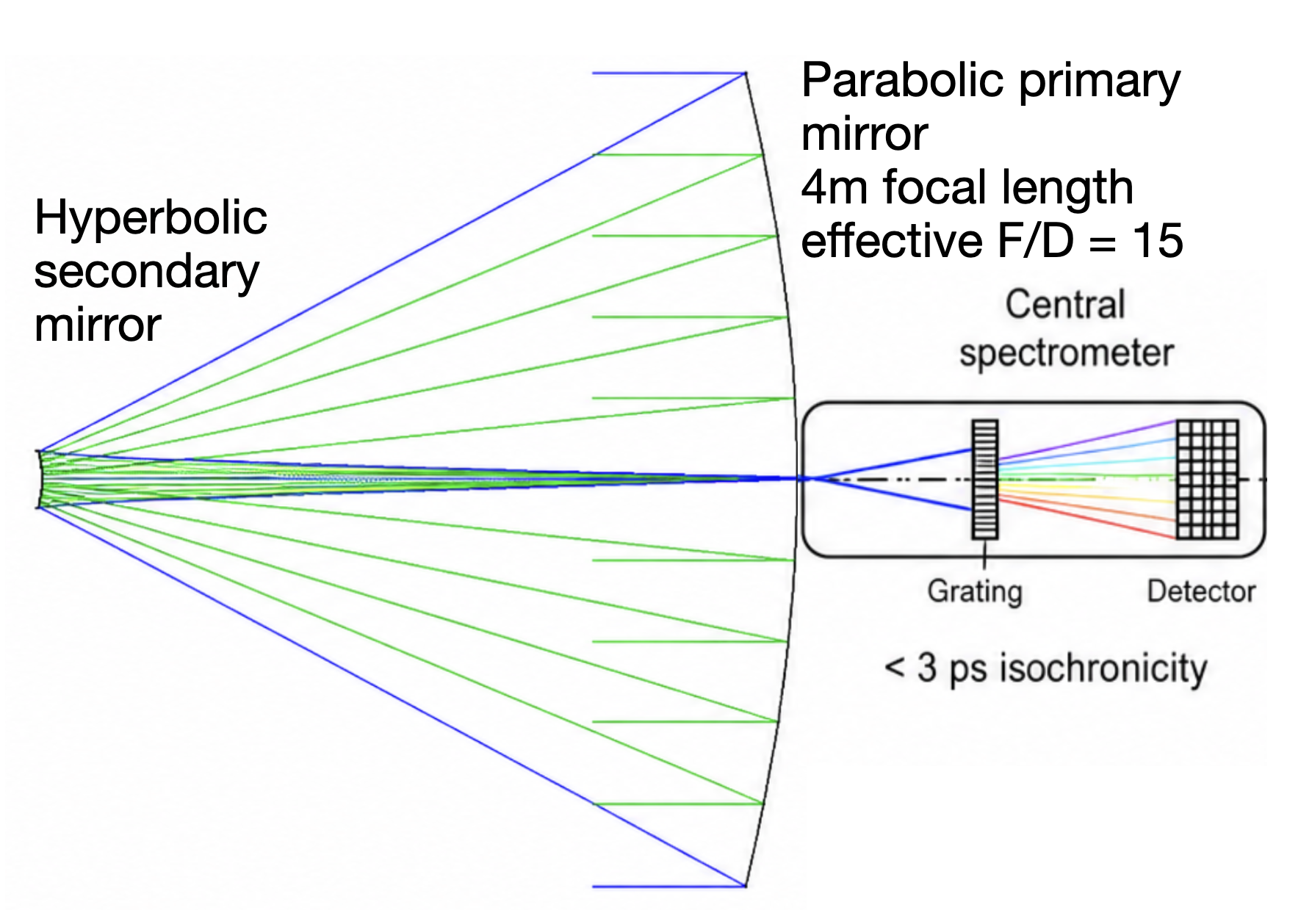}
    \caption{Cassegrain optics with fibre-free spectrograph feed. }
    \label{fig:cassegrain_optics}
    \end{subfigure}
    \caption{Primary-mirror and Cassegrain optical layout. (a) Segmented
        primary-mirror layout used for the current EON-SII collector design. The
        baseline uses round mirrors rather than hexagonal panels for cost
        reasons. (b) Optical layout optimized for photon concentration and
        stable spectrograph injection rather than diffraction-limited imaging.}
    \label{fig:mirrors}
\end{figure}

The current dish layout contains 18 off-axis parabolic primary panels with nominal
diameters near 0.8 m (Figs \ref{fig:telescope_design_concept} and
\ref{fig:round_mirror_design}). Each panel is mounted on a three-point active
support that adjusts its pointing direction. The actuator design range is
10 mm, with a target positioning accuracy of approximately $1\,\mu$m. Two mirror
technologies are currently being investigated: mould-free slumped glass
\citep{falk2025freeform} and diamond-machined aluminium. Slumped-glass mirrors
with a thickness of 2 cm are attractive because they can be reproduced
efficiently, but their forming accuracy, coating lifetime and final cost require
prototype qualification. Diamond-milled aluminium mirrors provide a contingency
option that increases capital cost but keeps production and design iteration
under direct engineering control.

The azimuth and elevation axes use paired preloaded drives to suppress backlash.
The drive system is specified to provide a pointing accuracy of about
3--5 arcsec. After model correction by the active mirror-control system, the
focal-spot pointing stability should be better than 0.3 arcsec, providing margin
relative to the 3-arcsec spectrograph-injection footprint. A separate stellar
guide camera measures the field position, while the active-optics loop corrects
the relative alignment of the mirror segments and the spectrograph entrance.

\begin{figure}
\centering
\includegraphics[width=\columnwidth]{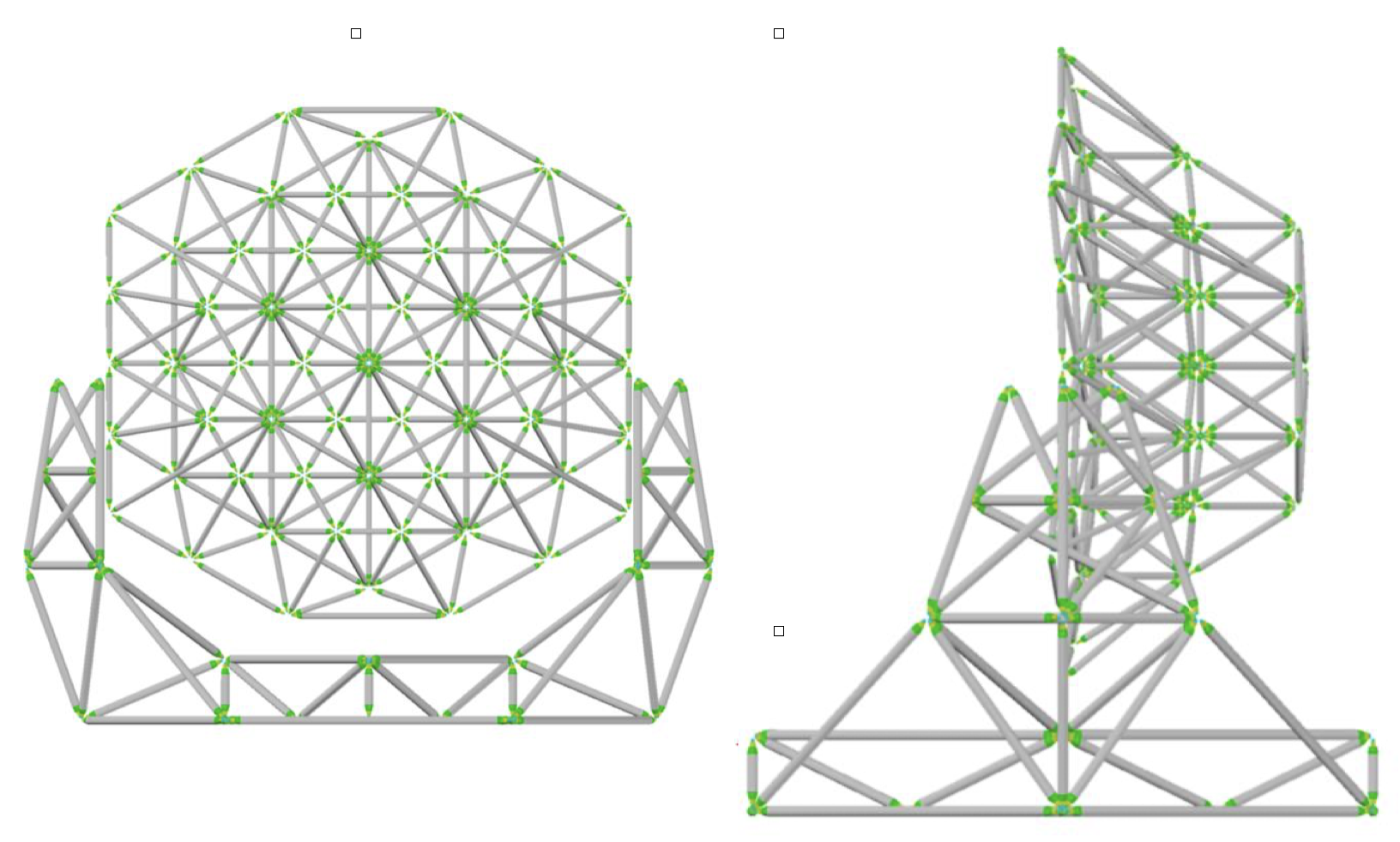}
\caption{MERO-TSK space-frame concept for the telescope structure.}
\label{fig:mero_baseline}
\end{figure}

\subsection{Transport and baseline reconfiguration}

The telescope is mounted on a road-transportable trailer. Four deployable feet
transfer loads to massive concrete blocks and provide a stable base in operation. The mast
and dish are folded or removed for transport, reducing the shipping envelope and
allowing both telescopes to be assembled and commissioned near the institute
before observatory deployment. Figure \ref{fig:operation_concept} shows an artists
impression of the operational concept.

Transportability is a scientific feature because the useful visibility slope
depends on target diameter. Short baselines validate the instrument on large,
bright calibrators; kilometre baselines address white dwarfs. A reconfiguration
changes the projected-baseline track without modifying the detector or
correlator. It also permits a staged commissioning sequence in which systematic
errors are isolated before the lowest-visibility observations.

\subsection{Cassegrain optical layout and active mirror control}

The chosen telescope optical system is a Cassegrain design with an effective
focal ratio of $F/D=15$. The focal length of the primary mirror is f=4~m.
A hyperbolic secondary reflects the beam back 
to the mechanically stiff dish centre, where the spectrograph and detector 
can be supported without a heavy focal-plane camera.
The focal length maps 3 arcsec to approximately 0.9 mm. This scale is large
enough for a practical entrance diaphragm.

Optical quality is specified by encircled energy rather than diffraction-limited
wavefront error. At least 90~per cent of the on-axis light must remain within the
3-arcsec entrance footprint over the permitted elevation and wind envelope.
That requirement is stringent enough to control background and spectrograph
loss, while being compatible with relatively low-cost segments and a lightweight
structure. It is also directly measurable with the acquisition camera. The
mirror-reflectivity assumptions follow experience from MAGIC mirror
characterization \citep{magic_reflectivity}.

The secondary mirror is mounted on fast piezo actuators with a 100 micron range to
permanently refocus the light at the entrance diaphragm of the spectrometer
with a pointing precision of <0.3~arcsec.
A small fraction of the light is sent to a camera via a beamsplitter, which 
estimates the focal-spot position
and shape. A closed-loop secondary stage with a target update rate near
100 Hz corrects pointing jitter and low-order dish motion. This system can
dynamically correct for drive control imperfections and wind oscillations of the
structure. Slower actuators
at the primary mirror segments permanently refocus the panels as gravity and 
temperature change. This is done by mounting CMOS cameras at each panel that 
see LEDs mounted at the frame holder of the secondary mirror. This allows
a permanent feedback loop, refocusing the mirrors for good PSF.
Separating fast beam steering from slow PSF control  separates the control
loop frequencies of the primary and the secondary panels.

\begin{figure}
\centering
\includegraphics[width=\columnwidth]{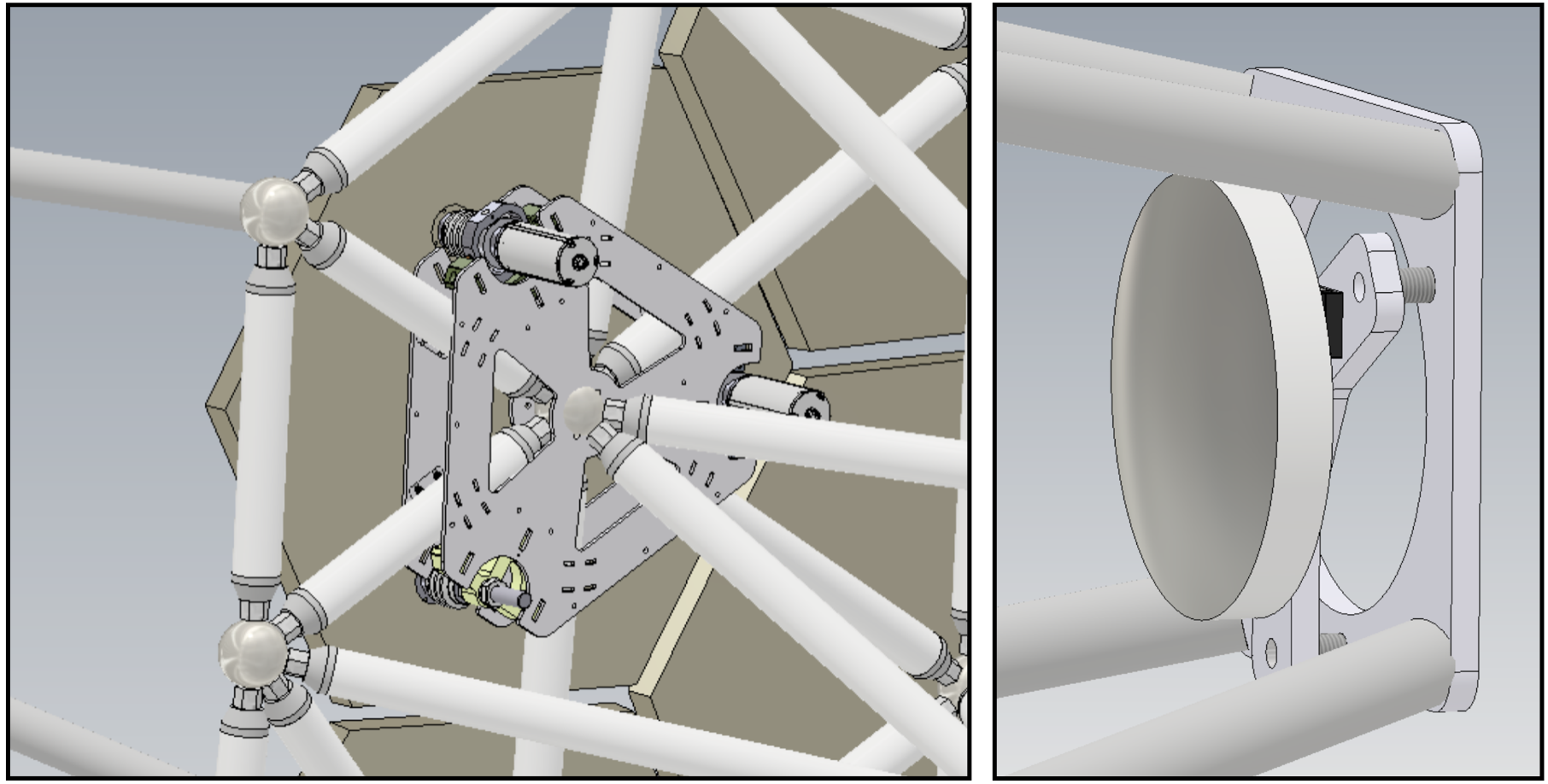}
\caption{Mirror-support and actuator concept for the segmented primary mirror.
The active support provide tip and tilt control of individual panels.}
\label{fig:mirror_fixation_concept}
\end{figure}

\begin{figure}
\centering
\includegraphics[width=\columnwidth]{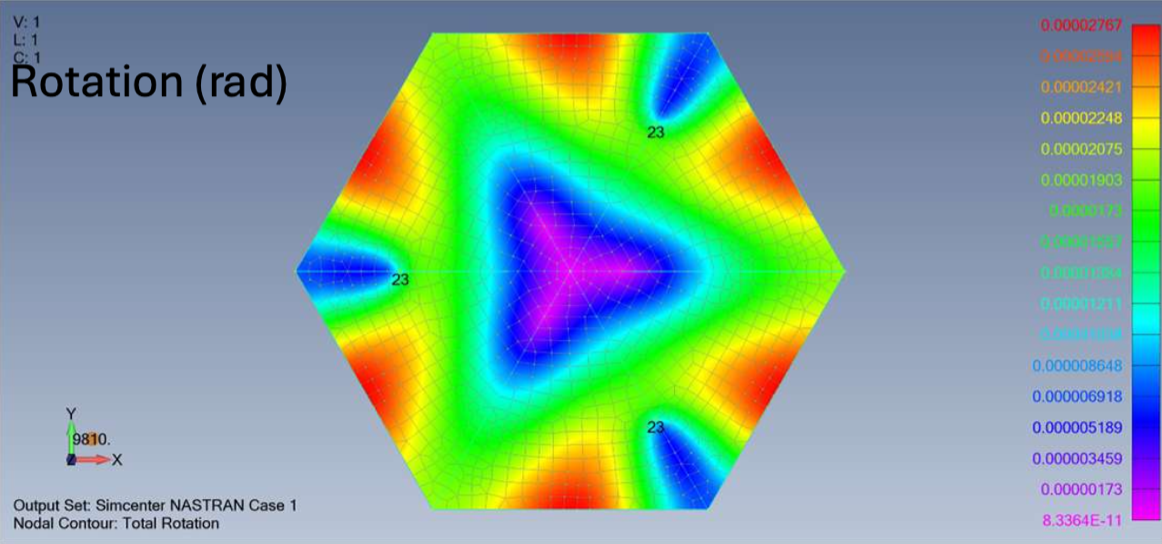}
\caption{Finite-element validation of a representative glass mirror panel
of 2 cm thickness.}
\label{fig:mirror_fem}
\end{figure}


\section{Spectrograph, detector and correlation architecture}
\label{sec:instrument}

\subsection{Fibre-free spectrograph}

\begin{figure}
    \centering
    \includegraphics[width=\columnwidth]{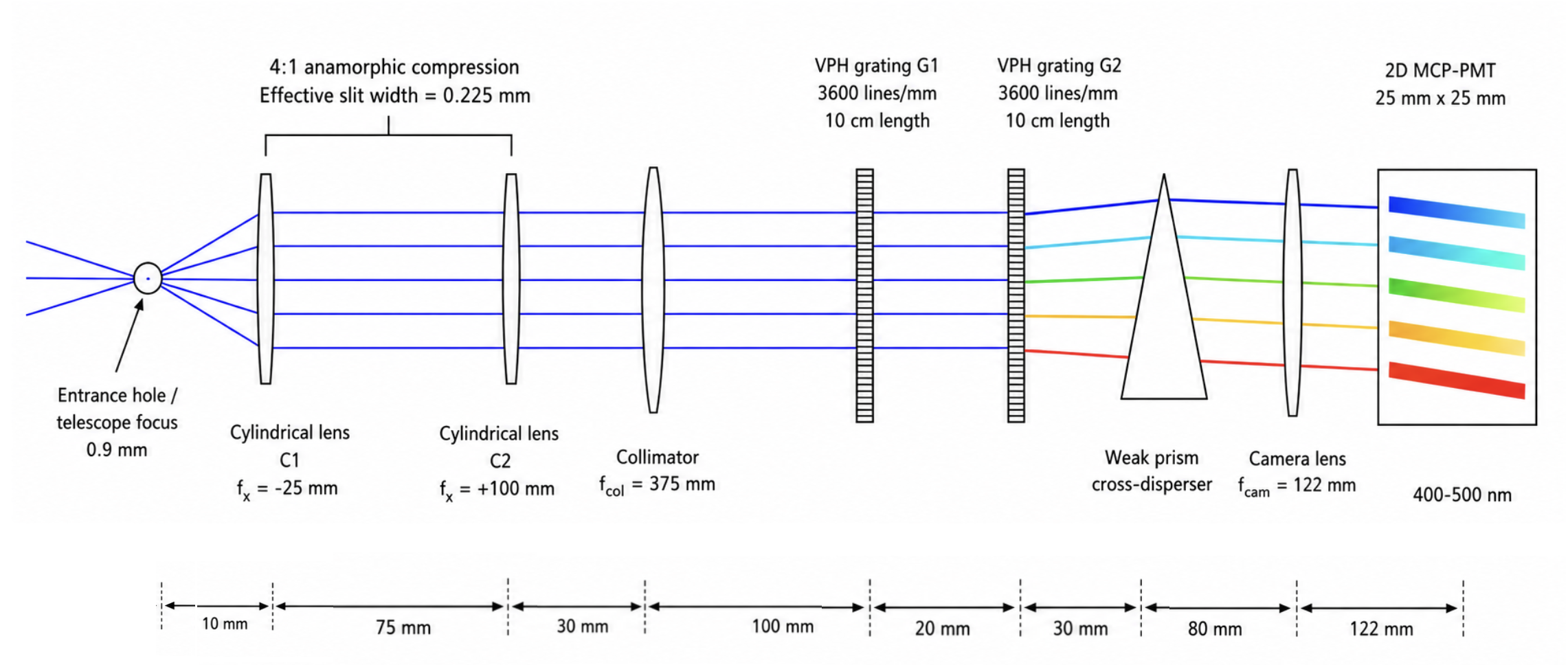}
    \caption{Fibre-free spectrograph concept. The anamorphic optics compresses
    the focal spot in one dimension, volume-phase gratings provide the principal
    dispersion, and a weak prism separates spectral rows on a two-dimensional
    photon-counting sensor.}
    \label{fig:spectrograph}
\end{figure}

Conventional fibre injection would add coupling loss, focal-ratio degradation
and wavelength-dependent transmission. This concept instead places the spectrograph at
the Cassegrain return focus and injects the focal spot directly through an
entrance diaphragm. An anamorphic optic compresses the approximately
0.9-mm spot in the dispersion direction to form a pseudo-slit. Two high-density
volume-phase gratings and a weak prism cross-disperser then map the 400--550 nm
band onto a two-dimensional sensor (Fig. \ref{fig:spectrograph}).

The reference resolving power is $R\simeq7000$--8000. Across the conservative
400--500 nm performance interval this gives of order $1000$ usable resolution
elements after sampling and edge losses. The requirement is not simply nominal
resolving power: the line-spread function must retain correlation contrast,
adjacent channels must be sufficiently independent, and total throughput after
the primary must exceed 60\%. Those quantities are measured by scanning
a narrow-line source across the full focal surface and deriving a channel
response matrix.

\subsection{Rejection of Sirius A}

Sirius B is separated from Sirius A by approximately 11 arcsec and is about
9000 times fainter in the visual band \citep{bond2017sirius}. If a fraction
$f_{A\rightarrow B}$ of the Sirius A photon rate enters the B extraction, the
contaminating rate is
\begin{equation}
R_{A\rightarrow B}
=f_{A\rightarrow B}R_A
\simeq9000f_{A\rightarrow B}R_B.
\label{eq:sirius_contamination}
\end{equation}

Keeping this term below 10~\% of the Sirius B rate requires
$f_{A\rightarrow B}\lesssim1.1\times10^{-5}$. The reference design combines a
focal-plane occulting mask centred on Sirius A, a reimaged pupil with a Lyot
stop, and the final entrance aperture centred on Sirius B.

The contrast requirement is one of the most demanding parts of the telescope. It
couples mirror scatter, panel-edge diffraction, atmospheric seeing, guiding,
occulting-mask alignment and spectrograph stray light.  
If the Sirius A requirement is not initially met, 40 Eridani B
and isolated stars of comparable magnitude provide commissioning targets while
the suppression system is improved later on.

\subsection{Photon sensors and front-end electronics}

\begin{figure}
\centering
\includegraphics[width=0.9\columnwidth]{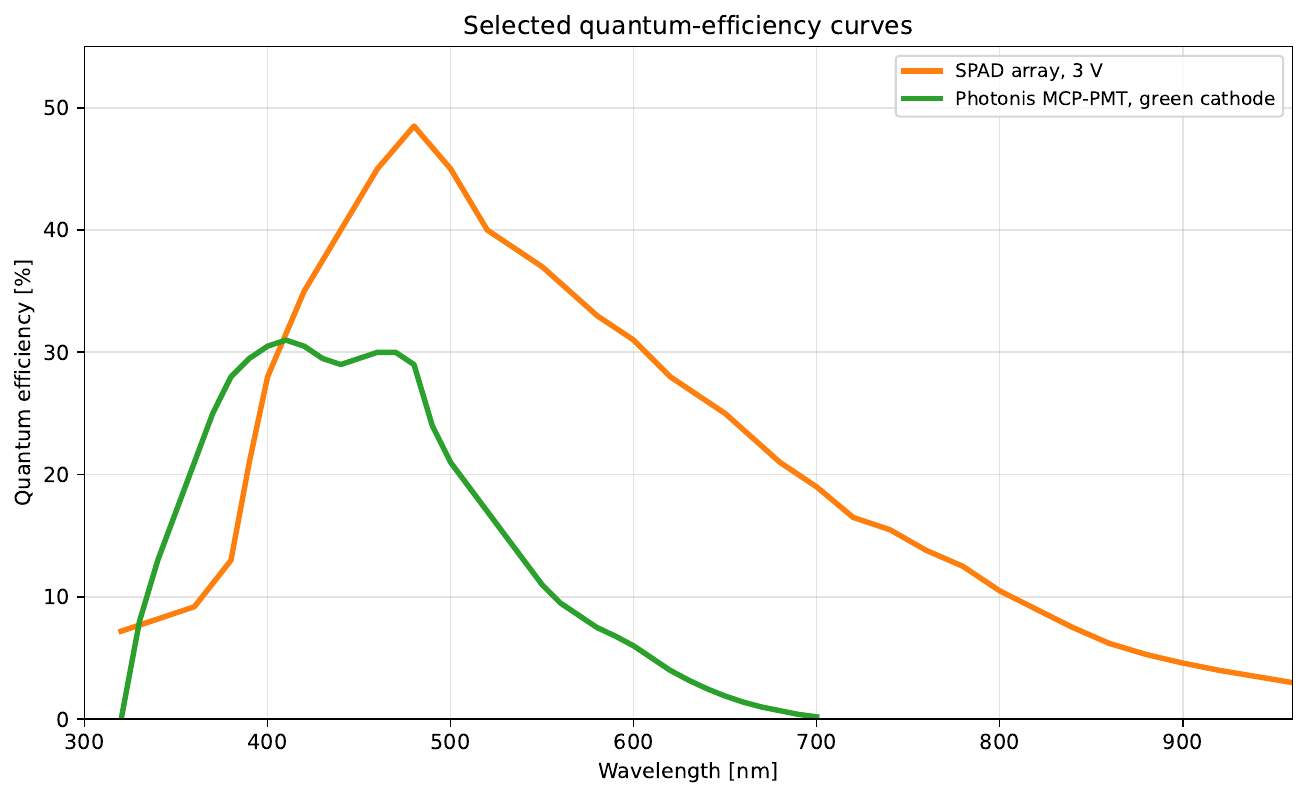}
\caption{Quantum-efficiency comparison of the Photonis FT18 MCP-PMT and the
QUASAR SPAD array used in the B2 detector trade. The two detector cases are
kept separate throughout the projected sensitivity analysis.}
\label{fig:detector_qe}
\end{figure}

The current design concept combines a 32x32 two-dimensional channel 
layout with fast single-photon response and separates
the initial design from the availability schedule of a custom SPAD array
(developed for the QUASAR project \citep{walter2025quasar}; R. Walter,
private communication, 2026). The SPAD array provides significantly faster timing
and higher QE. See Fig. \ref{fig:detector_qe}.

\begin{figure}
\centering
\includegraphics[width=0.9\columnwidth]{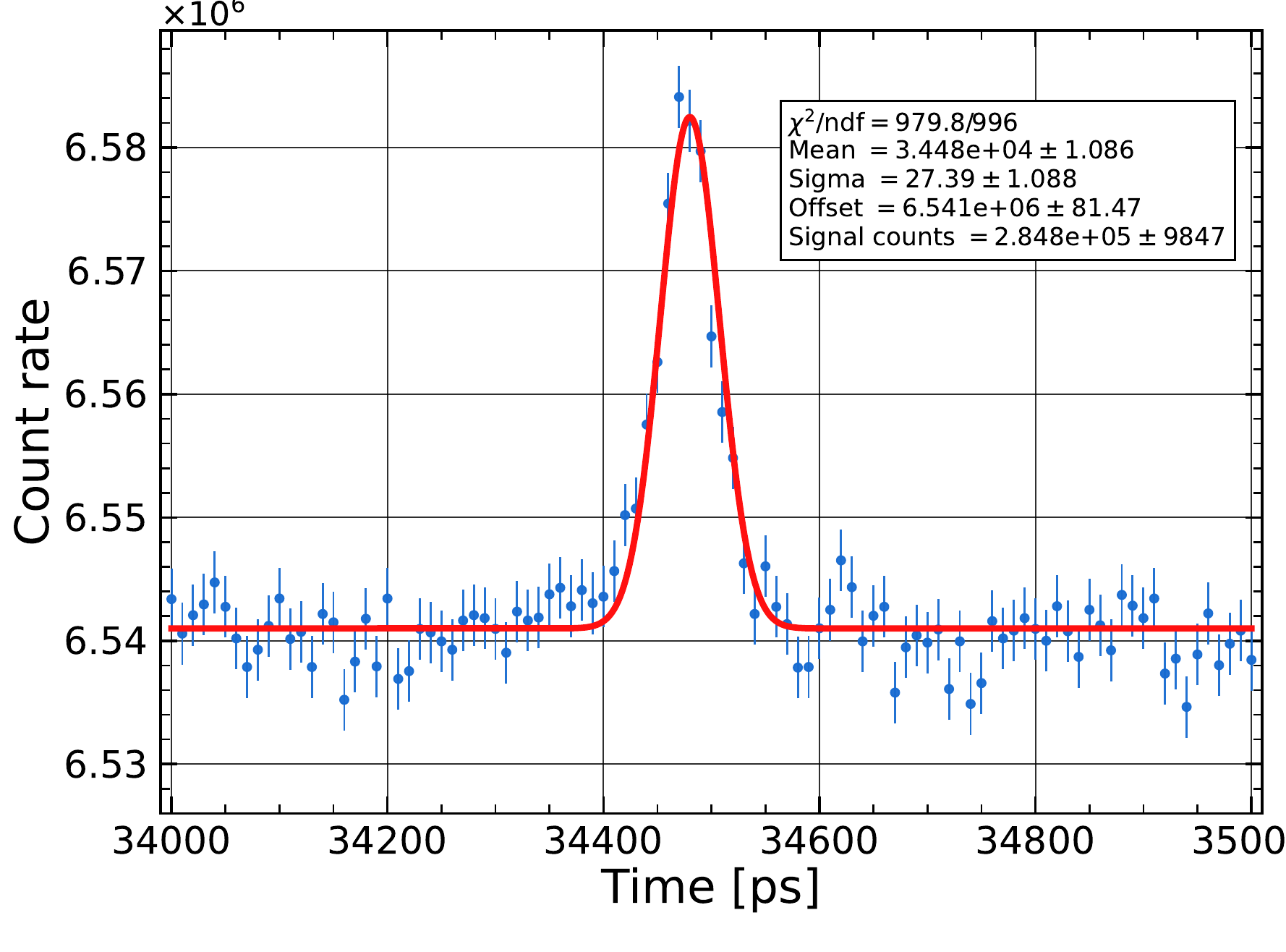}
\caption{Laboratory  zero-baseline  correlation peak (via beamsplitter)
of two Photonis FT18/Lidar MCP-PMT sensors with fitted Gaussian width
$\sigma=27.4\pm1.1$ ps. This measurement validates the timing scale of the
readout chain. We used for this measurement the Time Tagger X from Swabian 
Instruments.}
\label{fig:correlation_peak}
\end{figure}

A laboratory measurement of the Photonis FT18/Lidar device gave 32.4 ps FWHM
transit-time spread. A separate two-stream HBT test produced a fitted Gaussian
correlation width $\sigma=27.4\pm1.1$ ps, or approximately 64.5 ps FWHM, with a significance near 29 (Fig. \ref{fig:correlation_peak}). These measurements
validate the timing scale for individual components.

\begin{figure}
\centering
\includegraphics[width=\columnwidth]{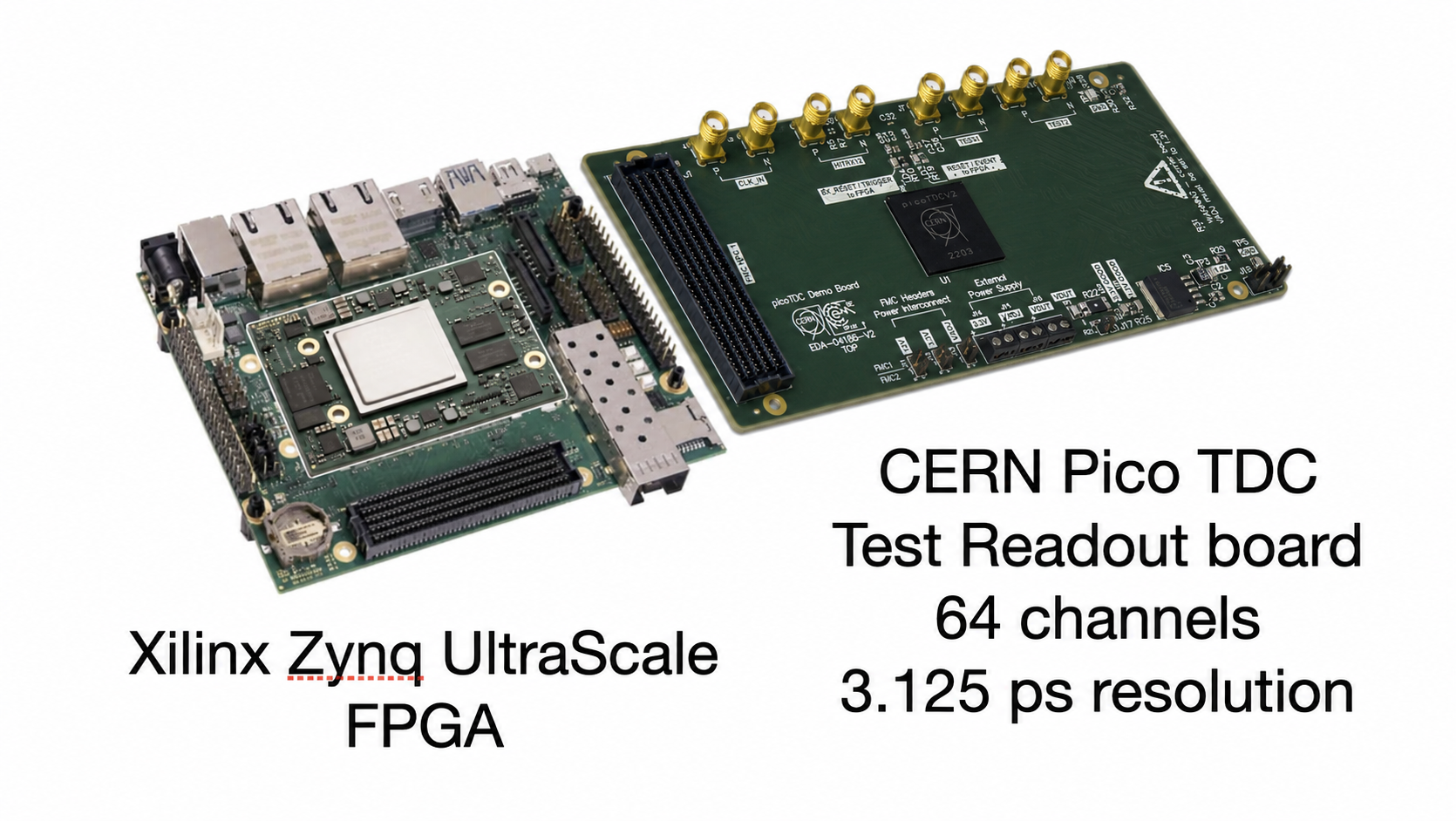}
\caption{64-channel picoTDC test board and Xylinx Zynq UltraScale FPGA 
carrier to evaluate picosecond time digitisation for the scalable 
EON-SII readout.}
\label{fig:picotdc_testboard}
\end{figure}

The upgrade path uses a two-dimensional SPAD array with higher photon-detection
efficiency and much better timing. Digital photon sensors are
particularly attractive for eliminating analogue gain variation and enabling
per-pixel masking, but optical fill factor, cross-talk, afterpulsing, dead time
and sustained bright-background operation must all be included in the final
trade \citep{liu2024sigeSpadArray,mirzoyan2026spad}. The paper therefore reports MCP-PMT and SPAD
performance cases separately.

\subsection{Time digitisation, transport and real-time correlation}

The basis of the TDC readout design is the 64-channel CERN picoTDC chip, which provides
a measured time resolution below 3 ps RMS and low-cost scaling to 1024
channels. A prototype board has been built and is being tested.

Sixteen  devices provide 1024 timestamp channels with
nominal 3.125-ps binning \citep{altruda2023}. Each layer includes programmable
front-end discrimination, channel calibration and an FPGA for event framing.
The estimated cost per channel is <~20~Euro/channel.
For Sirius B, the design model gives approximately 5-10 kHz per spectral channel,
or about 7 MHz per telescope over 1000 channels. Bright sources can increase the
trigger rate by orders of magnitude. Controlled mirror panel defocus and channel 
masking reduce the data rate to a manageable quantity of at most 1~GHz per
telescope. The limit for one channel shall be 100~MHz to keep the possibility
to observe single spectral lines.

Relative time transfer is calibrated in the same signal chain as the photons.
A fast optical pulse source, positioned at the center readout location between
both telescopes, feeds one reference channel at both telescopes
simultaneously and goes back to the centre to be tagged there. The measured
reference timestamps track clock phase and slowly 
varying cable or electronics delay. Geometric delays are calculated from the 
measured telescope coordinates in 3D, the surveyed baseline, source direction 
and time, then refined with bright star calibrators of known coordinates. 
The precision is demanding: it must be less than about 5 ps, but it is
manageable and calibratable.

Optical QSFP links transmit timestamp blocks to a central correlator. A staged
FPGA pipeline applies baseline delay correction, channel mapping, channel delay calibration,  fine delay and correlation,
then accumulates correlation histograms in time bins of around 0.1~sec and also
measures in real time the zero baseline correlation with itself. The design envelope is up
to $5\times10^9$ timestamps s$^{-1}$ per link. Raw or decimated diagnostic
streams are retained for offline validation, but routine observation cannot
depend on storing every event. A second software correlator processes selected
intervals and serves as an independent check on firmware results.

\section{End-to-end performance model}
\label{sec:performance}

\subsection{Monte Carlo study of telescope design parameters}

The telescope design parameters were derived from the requirements imposed by
the science goals described above \cite{Schweizer2026Simulation}. These goals
include observations of compact objects with magnitudes up to $V=10.7$ and
angular diameters down to $15$--$30~\mu$as. A simulation study indicates that
the following minimum instrumental parameters are required: a mirror area of
$9~\mathrm{m^2}$, a mirror reflectivity of 80\%, an optical throughput of 60\%
from the mirror to the sensor, a sensor timing resolution of 12--30 ps FWHM,
and reconfigurable baselines in the range 1.5--3 km.

The EON-SII performance model follows individual detected photons from the source
through atmosphere, optics, detector and correlation, following the general
approach of \citet{rou2013}. For a source spectrum $F_\lambda$, the expected
rate in spectral channel $i$ is constructed from the collecting area,
zenith-dependent atmospheric transmission, primary-mirror reflectivity,
spectrograph throughput, detector photon-detection efficiency, polarization
splitting and the channel line-spread function. Atmospheric transmission and
airmass terms follow standard extinction and airmass prescriptions
\citep{burki1995,kasten1965,kastenyoung1989}. Independent contributions from
night-sky background, dark counts and contaminating sources are added before
detector dead time and timing jitter are applied.

\subsection{Sirius B reference case and general sensitivity}

\begin{figure}
\centering
\includegraphics[width=\columnwidth]{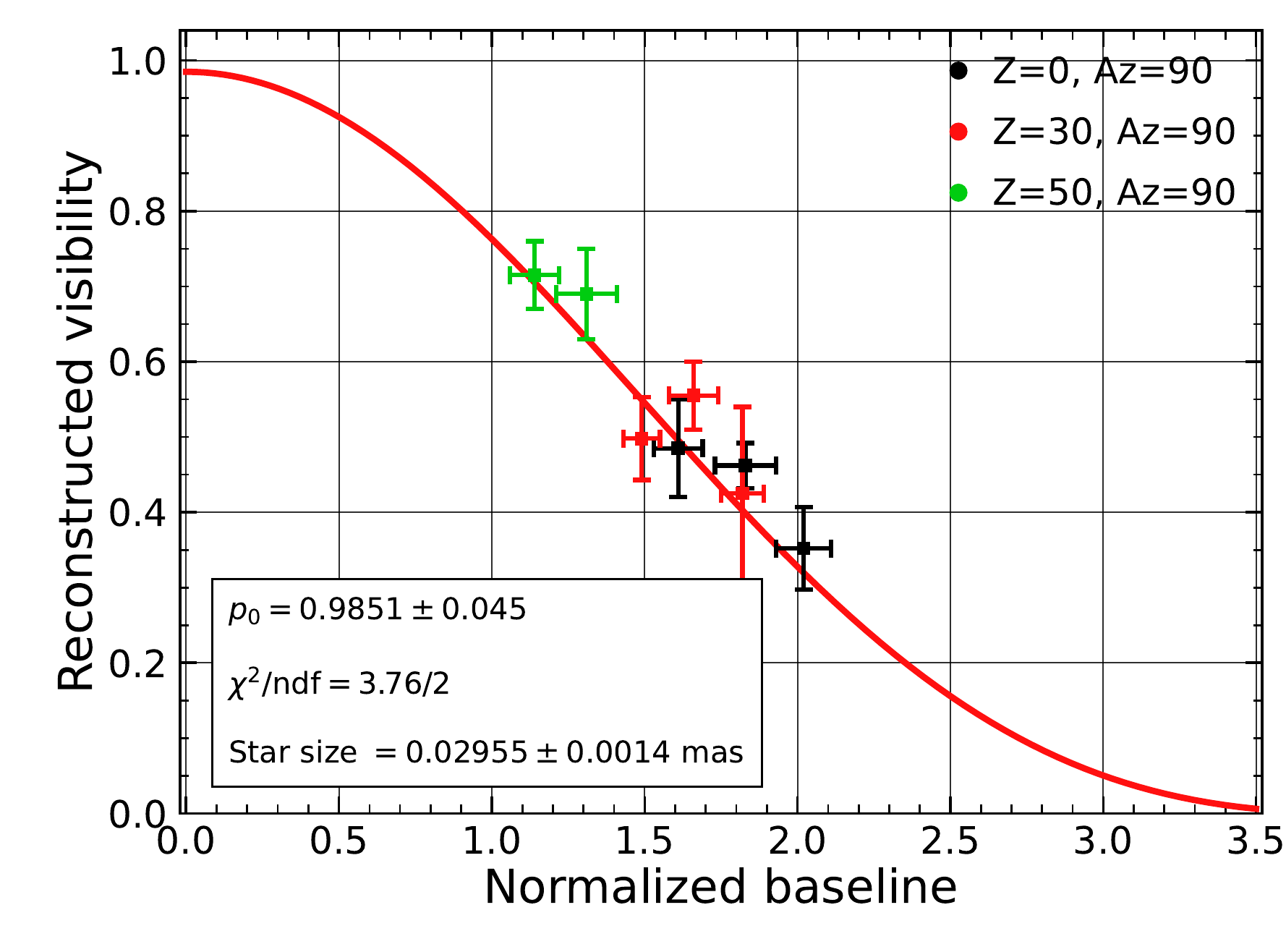}
\caption{Simulated Sirius B visibility data set at three zenith angles and a
uniform-disc fit. The horizontal axis is the baseline normalized to the
simulation reference scale. This is a simulation product, not on-sky EON-SII
data.}
\label{fig:sirius_visibility}
\end{figure}

At a nominal baseline of 1.5 km, Earth rotation provides several projected
baseline samples across the visibility slope of Sirius B. Figure
\ref{fig:sirius_visibility} shows a Monte Carlo simulation of the visibility
reconstruction from a 10 h observation of Sirius B, combining measured
visibility values from 1000 spectral channels at three different zenith angles.
Given the visibility function, the angular diameter of Sirius B can be
reconstructed. The simulation assumed an angular diameter of $0.030~\mathrm{mas}$
and, for a Photonis MCP-PMT sensor, recovered
$0.02955\pm0.0014~\mathrm{mas}$. The exact recovered value varies between
Monte Carlo realizations.

\begin{figure}
\centering
\includegraphics[width=\columnwidth]{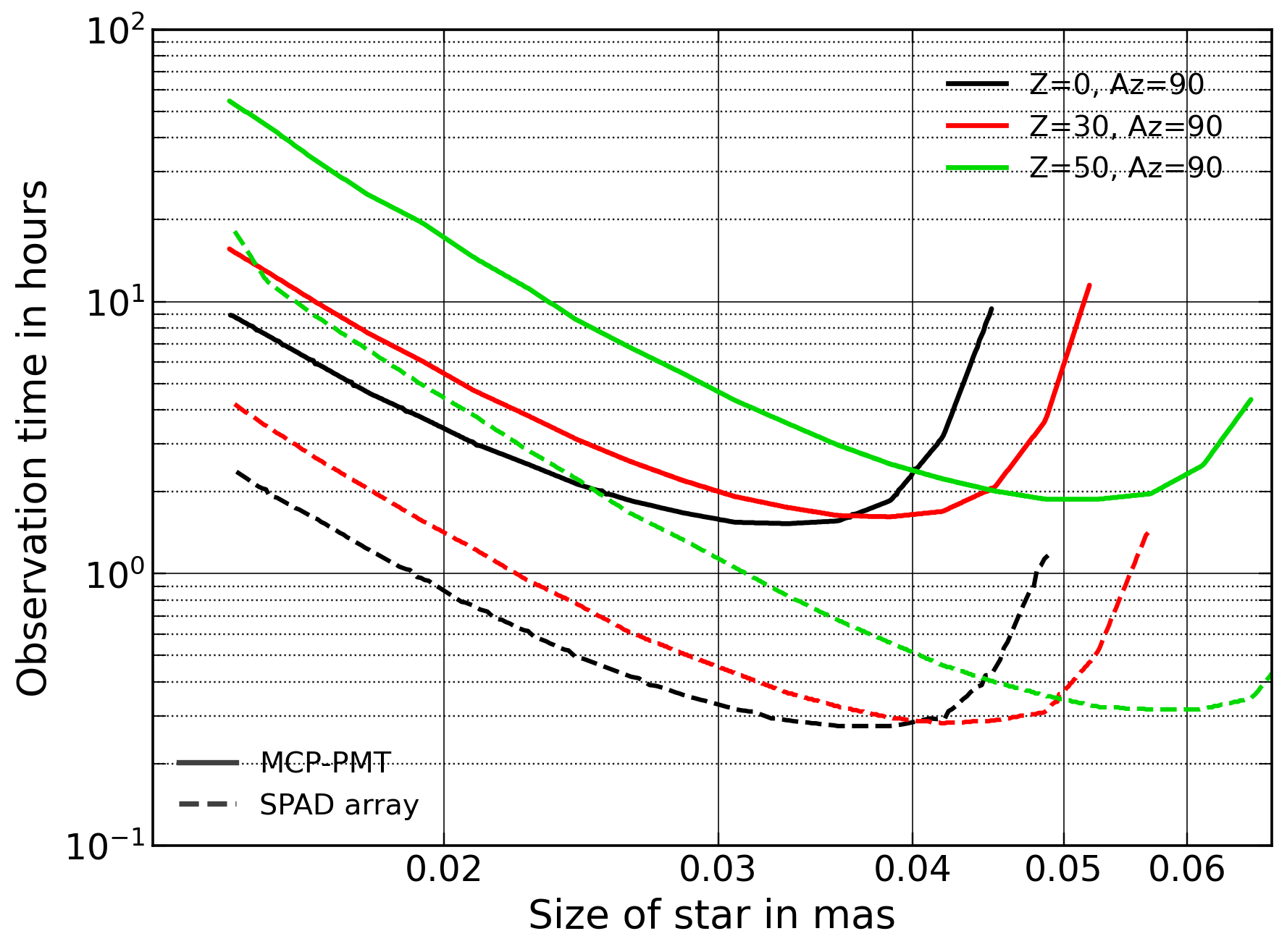}
\caption{Predicted time for the reference significance as a function of angular
diameter for the MCP-PMT and SPAD sensors; colours denote the zenith angle pointing,change in effective baseline and atmospheric extinction. The
curves are photon-level model predictions based on throughput, reflectivity, timing
and further assumptions listed in the text.}
\label{fig:sensitivity_comparison}
\end{figure}

Figure \ref{fig:sensitivity_comparison} compares the predicted sensitivities of
an MCP-PMT sensor and a SPAD-array sensor for a 1.5 km baseline.

The predicted integration time required to reach 10~\% diameter precision
is approximately 1.5 h for the conservative MCP-PMT case and 20 min for the
SPAD case at favourable zenith angle. Because visibility depends non-linearly
on angular diameter, the optimal integration time for a fixed baseline depends
on source size. Stars that are only marginally resolved provide little diameter
leverage, whereas stars observed beyond the first visibility lobe provide little
correlated signal.

Baseline reconfiguration is therefore required to observe different targets
near their optimal baseline regime.

Table \ref{tab:target_times} summarizes representative validation targets. The values
are photon-statistical predictions from the source model and are rounded
to avoid implying precision beyond the input assumptions. The Sirius B entries
for 5 and 2~\% diameter precision use the approximate
$T\propto\sigma_\theta^{-2}$ scaling from the 10~\% cases. Real
observations may reach a systematic floor set by transfer-function
calibration, residual Sirius A light, baseline knowledge and wavelength
calibration.

\begin{table*}
\centering
\caption{Projected integration times and baseline regimes for representative
white-dwarf validation cases. Times are photon-statistical predictions from the
end-to-end model and are used only after the corresponding systematic gates have
been passed. Dashes identify cases not retained as practical baseline
predictions.}
\label{tab:target_times}
\begin{tabularx}{\textwidth}{@{}lcccl@{}}
\toprule
Target & $m_V$ & Diameter & Baseline regime & $T_{10}/T_{5}/T_{2}$ \\
\midrule
Sirius B & 8.44 & 28.5 $\mu$as & 1.5--2.0 km & MCP-PMT: 1.5/6/37.5 h; SPAD: 0.33/1.3/8.2 h \\
40 Eridani B & 9.52 & 24.3 $\mu$as & 1.8--2.5 km & MCP-PMT: 10/40/-- h; SPAD: 2/8/50 h \\
Procyon B & 10.70 & 32.6 $\mu$as & 1.4--2.0 km & SPAD: 25-30/100/-- h \\
Bright CV or nova-like source & 8--11 & 20--500 $\mu$as & 0.5--3.0 km & MCP-PMT: $\gtrsim1$ h; SPAD: 0.2--30 h \\
\bottomrule
\end{tabularx}
\end{table*}

\subsection{From statistical precision to an angular diameter}

The diameter fit uses all channel and time-bin measurements rather than a single
averaged visibility. For measured  values $y_j=|V_j|^{2}$ with covariance $C$, the design
analysis minimizes the $\chi^2$:

\begin{equation}
\chi^2(\theta,\boldsymbol{a})
=
\left[\boldsymbol{y}-\boldsymbol{m}(\theta,\boldsymbol{a})\right]^{\rm T}
C^{-1}
\left[\boldsymbol{y}-\boldsymbol{m}(\theta,\boldsymbol{a})\right],
\label{eq:diameter_fit}
\end{equation}

where $\boldsymbol{m}(\theta,\boldsymbol{a})$ is the model function, depending
on the angular diameter and $\boldsymbol{a}$ contains transfer-function, 
background and limb-darkening nuisance parameters. Channel-to-channel 
correlations in $C$ result from a limited resolution power of the spectrometer 
and are measured with off-source and zero-baseline data. With the current
spectrometer resolution we expect a cross correlation between 5-7\%.
The requirement on electronic cross talk is set to <2\%.
As a fact: treating correlated spectral samples as independent would 
overstate the spectral channel number gain. (In the MC simulation study \cite{Schweizer2026Simulation} this has been taken into account within the 
visibility-efficiency $\eta$ parameter).

For the initial white-dwarf measurements, both uniform-disc and atmosphere-based
limb-darkened models will be fitted. The instrument first establishes a robust
angular scale; interpretation as a photospheric radius then combines the
diameter with parallax and a model convention. A 10~\% result demonstrates
the method, 5~\% begins to constrain the white-dwarf mass--radius relation,
and a 2~\% measurement requires the complete systematic budget to remain
below the photon-statistical error.

\subsection{Accretion and ejecta observables}

For time-dependent sources the same likelihood is applied to the half-radius
$R_{1/2}(\lambda,t)$, inclination, position angle and line-to-continuum flux
fractions. A uniform accretion disc gives $R_{1/2}=R_{\rm UD}/\sqrt{2}$, but dwarf novae
and novae are fitted with wavelength-dependent brightness distributions because
continuum, Balmer and He II emission trace different radii, different regions
in the source. The principal dwarf
novae are SS Cyg and U Gem: their expected tidal-disc diameters are
60--64 $\mu$as and their optical half-light diameters are expected to lie near
20--60 $\mu$as. Inside-out and outside-in models, enhanced mass-transfer and
truncated-disc models predict different signs and speeds of
$dR_{1/2}/dt$, different line-to-continuum size ratios and different lags
between optical brightening, He II emergence and spatial expansion
\citep{horne1986,marshhorne1988,smith2006vwhyi,hameury2020,baptista2001,dubus2018cv,kimura2026variable}.
These traces can be measured quantitatively with EON-SII and shed light
on competing models.

Classical and recurrent novae add a larger-scale target-of-opportunity 
case in which repeated continuum and Balmer/He II visibilities measure 
photospheric and ejecta expansion \citep{schaefer2014nova}.

\section{Discussion}
\label{sec:discussion}

\subsection{What the EON-SII reference design changes}

The EON-SII design combines four gains that are difficult to retrofit
simultaneously to existing telescopes. The dedicated telescope design delivers a compact
and actively stabilized focal spot. Direct spectrograph injection avoids fibre
coupling loss. Approximately 1000 independent channels provide statistical
multiplexing. Picosecond sensors and time digitisation reduce accidental
coincidences. None of these terms alone produces the quoted sensitivity: The
projected magnitude reach is an  consequence of their combination.

This coupling also means that nominal component specifications can be
misleading. A 3.125-ps TDC bin does not yield a 3.125-ps instrument response when
the detector is tens of picoseconds wide. A resolving power of 8000 does not
produce 1000 independent correlation measurements if the line-spread function
or electronics couples neighbouring pixels. A 4-m outer diameter does not
provide a filled 12.6-$\mathrm{m^2}$ aperture after segmentation. The paper uses
the effective measured quantities---area, throughput and correlation width---in the simulation study and final sensitivity budget\cite{Schweizer2026Simulation}.

\subsection{Initial science cases}

The primary validation case is a direct angular-diameter measurement of
Sirius B. Its accurately determined orbit and distance make it an unusually
valuable test of white-dwarf structure \citep{bond2017sirius}. Combining an EON-SII
angular diameter with parallax gives a physical radius. Combining that radius
with a dynamical mass places the star directly on the mass--radius relation of
electron-degenerate matter \citep{chandrasekhar1931,hamada1961}. The nearby
white dwarfs 40 Eridani B and Procyon B (if the simulated sensitivity is achieved) 
extend the method to independent masses, temperatures, atmospheric compositions and envelope structures \citep{bond2017eridani,bond2015procyon}. The target set is the complete bright, Tenerife-accessible non-accreting white-dwarf
sample emerging from the Gaia EDR3 white-dwarf catalogue and the 10-pc stellar
sample at the adopted magnitude limit (V<10.7)\citep{gentilefusillo2021,reyle2021}.

Bright fast rotators and binaries provide additional validation cases. Their
larger angular sizes allow shorter baselines and higher correlated signal, while
their non-circular visibility curves test orientation and baseline geometry.
Persistent high-state cataclysmic variables, dwarf-nova outbursts and bright
novae provide a time-domain extension. Wavelength-dependent characteristic
radii can distinguish changes in accretion-disc temperature structure, and
repeated tracks can measure evolving asymmetry. These targets require
observation-specific simulations because brightness, spectral lines and angular
scale change during an event \citep{hameury2020,baptista2001,dubus2018cv}.
The Ritter--Kolb catalogue indicates that the bright CV target set expands
rapidly between $V\leq10$ and $V\leq11$, making triggered observations a practical
part of the target set
\citep{ritterkolb2003}.

\begin{table*}
\centering
\caption{Primary white-dwarf and cataclysmic-variable targets for EON-SII. Visual
magnitudes are planning values; CV magnitudes are quoted as observed ranges.
Photospheric diameters use
$\theta_{\rm diam}[\mu{\rm as}]=9300.9(R_\star/R_\odot)/d_{\rm pc}$. CV
extended scales denote the approximate
accretion-disc, binary-separation or propeller-flow diameter, because the white
dwarf itself is generally below 1 $\mu$as.}
\label{tab:eon_wd_cv_targets}
\small
\begin{tabularx}{\textwidth}{@{}>{\raggedright\arraybackslash}p{0.16\textwidth}>{\raggedright\arraybackslash}p{0.10\textwidth}>{\raggedright\arraybackslash}p{0.27\textwidth}X@{}}
\toprule
Target & $V$ & Angular diameter or scale & Type and EON-SII role \\
\midrule
Sirius B & 8.44 & 28.5 $\mu$as diameter & DA white dwarf; primary mass--radius target, with stray-light suppression from Sirius A as the decisive observing risk \\
40 Eridani B & 9.52 & 24.3 $\mu$as diameter & DA white dwarf in a nearby triple system; clean independent mass--radius test \\
Procyon B & 10.7 & 32.6 $\mu$as diameter & cool white dwarf companion; high-value but difficult contrast target because of Procyon A \\
SS Cyg & 8.3--12.2 & 0.7 $\mu$as white dwarf; order 140--160 $\mu$as disc diameter; order 350 $\mu$as binary diameter & bright dwarf nova; outburst and high-state disc-structure target \\
U Gem & 8.2--14.9 & 0.7 $\mu$as white dwarf; order 120--130 $\mu$as outburst-disc diameter; order 290 $\mu$as binary diameter & archetypal eclipsing dwarf nova; line-dependent disc-radius target \\
AE Aqr & 10.9--12 & 1.0 $\mu$as white dwarf; order 500 $\mu$as binary/propeller-flow scale & magnetic propeller CV; tests whether EON-SII can constrain non-disc mass-transfer geometry \\
\bottomrule
\end{tabularx}
\end{table*}

The table separates two different target classes. For white dwarfs, the goal is to measure the size of the star itself. For CVs and related systems, the goal is to constrain the structure of the surrounding accretion flow, such as the disc or bright emission regions. 

The three nearby white dwarfs have predicted
diameters of 25--33 $\mu$as and therefore define the direct mass--radius case
\citep{bond2017sirius,shipman1997fortyeri,provencal1997procyon}. The CV white
dwarfs themselves are too small for the two-telescope pathfinder, but the
optically variable discs, Roche-lobe scales and stream or propeller structures
can reach 100--500 $\mu$as in favourable high states
\citep{sion2018cvupdate,bitner2007sscyg,smak2001ugem,wynn1997aeaqr}.

\begin{table*}
\centering
\caption{Compact-object and massive-binary targets for extended EON-SII
science cases. Visual magnitudes and angular diameters are approximate
planning values. Photospheric diameters use
$\theta_{\rm diam}[\mu{\rm as}]=9300.9(R_\star/R_\odot)/d_{\rm pc}$; line-disc
diameters refer to the characteristic H$\alpha$ or H$\beta$ emitting region
where listed.}
\label{tab:eon_compact_targets}
\small
\begin{tabularx}{\textwidth}{@{}>{\raggedright\arraybackslash}p{0.15\textwidth}>{\raggedright\arraybackslash}p{0.09\textwidth}>{\raggedright\arraybackslash}p{0.25\textwidth}X>{\raggedright\arraybackslash}p{0.17\textwidth}@{}}
\toprule
Target & $V$ & Angular diameter & Type & Microquasar status \\
\midrule
Cyg X-1 & 8.9 & 94 $\mu$as & O-supergiant high-mass X-ray binary with a black hole & secure, persistent \\
X Per & 6.7 & 75 $\mu$as & Be/X-ray binary with a neutron star & no \\
A 0535+26 / V725 Tau & 8.9--9.0 & 70 $\mu$as star; 214--223 $\mu$as H$\beta$ disc; 447--484 $\mu$as H$\alpha$ disc & Be/X-ray binary with a neutron star & no \\
HESS J0632+057 / MWC 148 & 9.1 & 48 $\mu$as star; order 820 $\mu$as line-disc scale & Be gamma-ray binary with a compact object & no \\
4U 2206+54 & 9.9 & 22 $\mu$as & peculiar O-star high-mass X-ray binary, neutron-star or magnetar candidate & no \\
LS I +61 303 & 10.6--10.8 & 24 $\mu$as & Be gamma-ray binary with an unsettled compact-object engine & debated, not secure \\
V4641 Sgr & 13.4--13.8 quiescent; 8.8 in major optical outburst & 8 $\mu$as & black-hole X-ray transient with a B-type donor & secure, target-of-opportunity only \\
Vela X-1 & 6.9 & 153 $\mu$as & B-supergiant high-mass X-ray binary with a neutron star & no \\
4U 1700-37 / HD 153919 & 6.5 & 107 $\mu$as & O-supergiant high-mass X-ray binary with a compact object & no \\
MWC 656 & 8.7--8.8 & 28 $\mu$as & Be binary; former black-hole candidate, now not treated as a microquasar & no \\
\bottomrule
\end{tabularx}
\end{table*}

Table \ref{tab:eon_compact_targets} gives a second set of science cases for the
instrument: bright compact-object binaries and gamma-ray binaries. 

These objects test whether the same optical system can measure stellar 
distortion, Be-disc size and wind-fed mass-transfer
geometry on 10--1000 $\mu$as scales. Cyg X-1 is the only optically bright,
persistent and secure microquasar in this list. Cyg X-1 is particularly
interesting since it is an established gamma-ray emitter up to energies >100~TeV
detected by LHAASO\citep{lhaaso2025microquasars}. V4641 Sgr is a confirmed
microquasar or microblazar, but its normal quiescent brightness makes it a
target-of-opportunity source rather than a regular observing case
\citep{salvesen2020v4641,revnivtsev2003v4641}. LS I +61 303 is retained as a
useful gamma-ray binary, while its interpretation as an accretion-powered
microquasar remains contested \citep{chernyakova2023lsi,papitto2012lsi}.
EON-SII measurements would provide spatial constraints on this question.

\subsection{Limitations of a two-telescope pathfinder}

Two telescopes provide one instantaneous baseline and hence one visibility
amplitude at each wavelength. Earth rotation increases the sampled track, and
transport changes its scale and orientation, but general image reconstruction
cannot be done with this system. The EON-SII system therefore will be evaluated on
calibrated visibilities, diameters, binary parameters and simple geometric
models. Reconstructed  images belong to a later many-telescope system.

The sensitivity predictions are  incomplete until the integrated instrument
is measured. The current model includes source spectrum, atmosphere, throughput,
quantum efficiency, background, timing, dead time and spectral response. The
dominant uncertainties are the perfection of the active mirror control
system in outdoor conditions, achieved resolution in dependence on
the quality of the optical focus and the Sirius A residuals after suppression.
Systematic visibility calibration, rather than raw photon statistics, is likely
to set the floor for the highest-precision diameter measurements.

Transportability introduces operational complications and additional alignment
effort. A moved telescope requires a new survey, pointing model,
active-optics calibration and delay verification. That means each baseline
will be exploited rather longer term before moving to the next telescope distance.
The scientific value of an optimized baseline must be
weighed against this reconfiguration cost. A practical observing strategy will
therefore use a small set of characterized pads rather than continuously varying
the telescope separation.

\subsection{Scaling to an imaging array}

An array of $N$ telescopes provides $N(N-1)/2$ simultaneous pair baselines. The
gain is primarily Fourier coverage and calibration redundancy. Sensitivity does
not simply scale with the number of pairs when baselines resolve different
spatial frequencies. Three or four telescopes already sample multiple baselines simultaneously, adding independent points in the uv plane and significantly improving constraints on fitted source models.

Tens of telescopes provide the dense $(u,v)$ sampling required for phase
retrieval and model-independent reconstruction
\citep{fienup1978fourierModulus,fienup1982phaseRetrieval,nunez2012,nunez2012imaging,nunez2012features,Schweizer2026Simulation}.

EON-SII's actively controlled, moderate-quality mirror panels are chosen for
cost-effective replication. The below-EUR-1-million cost constraint shifts the
design space toward repeated active collectors rather than scaled versions of
precision optical telescopes. The two-telescope system is therefore a test of
the collector, spectrograph and timing architecture before any three- or
four-telescope extension is considered.

\subsection{Comparison of small-telescope and Cherenkov telescope implementations}

Small-telescope experiments have demonstrated modern digital correlation and
stellar measurements, including observations of Sirius A with 0.25-m telescopes
\citep{mozdzen2025sirius}. Cherenkov arrays provide much larger collection area
and multiple baselines at existing sites. EON-SII occupies a different design point:
it gives up a wide air-shower field of view and nanosecond imaging camera in
exchange for compact on-axis injection, spectral parallelism, picosecond timing
and dedicated observing time. The decisive comparison is not mirror diameter
alone but the effective sensitivity in units of V-band magnitudes.


\section{Conclusions}
\label{sec:conclusions}

We have presented a complete reference design for the EON-SII two-telescope
system and traced its main engineering choices to the photon-correlation
measurement. The design uses two transportable 4-m, approximately
$9\,\mathrm{m^2}$ mirrors consisting of actively aligned segmented off-axis 
parabolic panels. It uses a Cassegrain optical system with an effective $F/D=15$
and a direct injection into a 400--550 nm spectrograph, as well as 
1000 effective channels with  picosecond time-tagging and a centralized
digital correlation.

Laboratory tests support the required timing scale at component level. A
candidate detector produced 27.4~ps RMS  HBT correlation signal and the picoTDC
architecture provides 3.125-ps nominal digitisation. Scaling this performance
uniformly to 1024 channels remains an integrated-system validation step.

The photon-level Monte Carlo model reproduces the integration-time scale of a
MAGIC/Adhara benchmark\cite{Schweizer2026Simulation} and predicts that the 
reference EON-SII configuration can
measure the angular diameter of Sirius B to 10~\% in approximately 1.5 h
with the conservative MCP-PMT detector or 20 min using a SPAD array. These values
are model predictions. The final result depends on achieved throughput,
spectral-channel independence, stable focal injection, long-duration timing and
suppression of Sirius A.

Four measurements define the experimental acceptance tests for the intended
sensitivity threshold: a stable 3-arcsec focal spot at the entrance
of the spectrometer on a moving
telescope, greater than 60~\% spectrograph throughput with approximately
1000 independent channels, end-to-end picosecond correlation at operational
rates, and Sirius A leakage at or below $10^{-5}$. These quantities are treated
as acceptance criteria rather than demonstrated system performance.

A successful two-element telescope system would establish calibrated microarcsecond
visibility measurements for compact stellar systems and validate a replicable
telescope readout architecture. Its broader role is to test whether massively
parallel spectral splitting and picosecond SII can be extended from present
bright-star demonstrations toward multi-aperture microarcsecond imaging arrays.


\section*{Acknowledgements}
The authors thank the engineers and technicians at the Max Planck Institute for Physics who contributed to the underlying component studies. The structural concept was developed with input from Josef Eder and MERO-TSK. The design also builds on experience gained within MAGIC, LST and the wider stellar intensity-interferometry community. AI systems have been used to improve English grammar and wording.

\section*{Author contributions}
Thomas Schweizer developed the reference design, the performance model, and the full simulation code, directed the component tests, and prepared the manuscript. Jürgen Besenrieder simulated the Cassegrain telescope optics. Josef Eder designed the telescope structure together with MERO-TSK and performed the FEM simulations of the mirror panels. Carina Haider performed the laboratory measurements. Razmik Mirzoyan contributed key expertise on telescope design and sensor performance. Olaf Reimann and Derek Strom provided design-improvement suggestions and actively supported the laboratory measurements. Roland Walter provided information on the SPAD-array development for the QuASAR project. All authors contributed to the scientific or technical development of the concept and reviewed the manuscript.

\section*{Conflict of interest}
The authors declare no conflict of interest.

\section*{Data availability}
The design parameters needed to reproduce the analytic estimates are given in this paper. The photon-level simulation inputs, derived numerical data underlying the projected-performance curves, and analysis scripts will be made available by the corresponding author on reasonable request.

\label{lastpage}

\end{document}